\documentclass[apj]{emulateapj}
\usepackage{graphicx}
\usepackage{epsfig}
\usepackage{float}
\providecommand{\e}[1]{\ensuremath{\times 10^{#1}}}

\newcommand{\noprint}[1]{}
\newcommand{\figsetstart}{{\bf Fig. Set} }
\newcommand{\figsetend}{}
\newcommand{\figsetgrpstart}{}
\newcommand{\figsetgrpend}{}
\newcommand{\figsetnum}[1]{{\bf #1.}}
\newcommand{\figsettitle}[1]{ {\bf #1} }
\newcommand{\figsetgrpnum}[1]{\noprint{#1}}
\newcommand{\figsetgrptitle}[1]{\noprint{#1}}
\newcommand{\figsetplot}[1]{\noprint{#1}}
\newcommand{\figsetgrpnote}[1]{\noprint{#1}}

\providecommand{\e}[1]{\ensuremath{\times 10^{#1}}}

\def\arcsec{$^{\prime\prime}$}

\shorttitle{SEDs of White Dwarfs in 47 Tuc}
\shortauthors{Woodley et al.}

\begin{document}

\title{The Spectral Energy Distributions of White Dwarfs in 47 Tucanae: The Distance to the Cluster\footnote{Based on observations with the NASA/ESA Hubble Space Telescope, obtained at the Space Telescope Science Institute, which is operated by the Association of Universities for Research in Astronomy, Inc., under NASA contract NAS 5-26555. These observations are associated with proposal GO-11677.}}

\author{K.~A.~Woodley\altaffilmark{1}, R.~Goldsbury\altaffilmark{1}, J.~S.~Kalirai\altaffilmark{2,3}, H.~B.~Richer\altaffilmark{1}, P.-E.~Tremblay\altaffilmark{4}, J.~Anderson\altaffilmark{2}, P.~Bergeron\altaffilmark{5}, A.~Dotter\altaffilmark{2}, L.~Esteves\altaffilmark{6}, G.~G.~Fahlman\altaffilmark{7}, B.~M.~S.~Hansen\altaffilmark{8}, J.~Heyl\altaffilmark{1}, J.~Hurley\altaffilmark{9}, R.~M.~Rich\altaffilmark{8}, M.~M.~Shara\altaffilmark{10}, P.~B.~Stetson\altaffilmark{7}}
\altaffiltext{1}{Department of Physics \& Astronomy, University of British Columbia, Vancouver, BC, Canada V6T 1Z1; kwoodley@phas.ubc.ca, rgoldsb@phas.ubc.ca, richer@astro.ubc.ca, heyl@phas.ubc.ca}
\altaffiltext{2}{Space Telescope Science Institute, 3700 San Martin Drive, Baltimore, MD, 21218; jkalirai@stsci.edu, jayander@stsci.edu, dotter@stsci.edu}
\altaffiltext{3}{Center for Astrophysical Sciences, Johns Hopkins University, Baltimore, MD, 21218}
\altaffiltext{4}{Zentrum f{\"u}r Astronomie der Universit{\"a}t Heidelberg, Landessternwarte, K{\"o}nigstuhl 12, 69117 Heidelberg, Germany; ptremblay@lsw.uni-heidelberg.de}
\altaffiltext{5}{D\'epartement de Physique, Universit\'e de Montr\'eal, C.P.~6128,
Succ.~Centre-Ville, Montr\'eal, Qu\'ebec H3C 3J7, Canada; bergeron@astro.umontreal.ca}
\altaffiltext{6}{Department of Physics, University of Guelph, Guelph, ON, Canada N1G 2W1; lesteves@uoguelph.ca}
\altaffiltext{7}{National Research Council, Herzberg Institute of Astrophysics, Victoria, BC, Canada V9E 2E7; greg.fahlman@nrc-cnrc.gc.ca, peter.stetson@nrc-cnrc.gc.ca}
\altaffiltext{8}{Division of Astronomy and Astrophysics, University of California at Los Angeles, Los Angeles, CA, 90095; hansen@astro.ucla.edu, rmr@astro.ucla.edu}
\altaffiltext{9}{Centre for Astrophysics \& Supercomputing, Swinburne University of Technology, Hawthorn, VIC 3122, Australia; jhurley@swin.edu.au}
\altaffiltext{10}{Department of Astrophysics, American Museum of Natural History, Central Park West and 79th Street, New York, NY 10024; mshara@amnh.org}

\begin{abstract}
We present a new distance determination to the Galactic globular cluster 47 Tucanae by fitting the spectral energy distributions of its white dwarfs to pure hydrogen atmosphere white dwarf models.  Our photometric dataset is obtained from a 121 orbit Hubble Space Telescope program using the Wide Field Camera~3 UVIS/IR channels, capturing F390W, F606W, F110W, and F160W images.  These images cover more than $60$ arcmin$^2$ and extend over a radial range of $5-13.7$ arcmin ($6.5-17.9$ pc) within the globular cluster. Using a likelihood analysis, we obtain a best fitting unreddened distance modulus of (m - M)$_{\rm{o}}=13.36\pm0.02\pm0.06$ corresponding to a distance of $4.70\pm0.04\pm0.13$ kpc, where the first error is random and the second is systematic.  We also search the white dwarf photometry for infrared excess in the F160W filter, indicative of debris disks or low mass companions, and find no convincing cases within our sample.

\end{abstract}


\section{Introduction}
\label{sec:intro}

Many astrophysical analyses involving globular clusters rely on well determined metallicity, reddening, and distance measurements.  In particular, the estimated ages of Galactic globular clusters can depend heavily on these assumed quantities, so their careful determination is necessarily important.  Ages of Galactic globular clusters can place strong constraints on the epoch of formation of our Galaxy, identify when cluster formation occurred with respect to the era of reionization in the Universe and the cosmic star formation peak, constrain the formation history of the Galaxy by dating the halo and bulge, as well as constrain $\Omega_M$ \citep{gratton03} when coupled with measurements of $H_{\rm{o}}$ by WMAP \citep{spergel03} and the HST Key Project \citep{freedman01}.  Most of the Galactic globular clusters have ages between 9-14 Gyr \citep[e.g.][]{marinfranch09,dotter10}, with most absolute ages commonly determined by fitting theoretical isochrones to the main sequence turnoff.  This, however, depends strongly on the assumed distance modulus to the cluster, which is the most significant contributor to the age uncertainty \citep{bolte95}.  Although less sensitive to distance, even the more recently employed age dating technique via the white dwarf cooling sequence \citep{hansen04,hansen07} requires a precise distance modulus.
\begin{deluxetable*}{lll} 
\tablecolumns{3}
\tabletypesize{\small} 
\tablecaption{Comparative Absolute Distance Estimates to 47 Tuc\label{tab:distance}}
\tablewidth{0pt}
\tablehead{
\colhead{$\rm{(m - M)_o}$\tablenotemark{a}} & \colhead{Method}&
\colhead{Reference}  \\
\colhead{ } &\colhead{ } &\colhead{ }  \\
}\startdata
$13.36\pm0.02\pm0.06$\tablenotemark{b} & WD Spectral Energy Distributions & This Paper \\
$13.23\pm0.08$ & Eclipsing Binary & \cite{thompson10}\\
$13.25$ & Main Sequence Fitting & \cite{bergbusch09} \\
$13.32\pm0.09$ & Tip of the Red Giant Branch & \cite{bono08}\\
$13.47\pm0.11$ & RR Lyrae & \cite{bono08}\\
$13.28\pm0.07$  &  Eclipsing Binary & \cite{kaluzny07} \\
$13.18\pm0.03\pm0.04$\tablenotemark{b} & Horizontal Branch Fitting & \cite{salaris07}\\
$13.02\pm0.19$& Cluster Kinematics & \cite{mclaughlin06}\\
$13.40\pm0.08$\tablenotemark{c,d} & Main Sequence Fitting &  \cite{gratton03} \\ 
$13.25+0.06-0.07$  & Main Sequence Fitting & \cite{percival02} \\
$13.21\pm0.04\pm0.07$\tablenotemark{b} & Main Sequence Fitting & \cite{grundahl02} \\
$13.15\pm0.14$ & WD Cooling Sequence Fitting & \cite{zoccali01} \\ 
$13.45\pm0.09$\tablenotemark{c} & Main Sequence Fitting & \cite{carretta00} \\
$13.32\pm0.20$ & Tip of the Red Giant Branch & \cite{ferraro00} \\ 
$13.29\pm0.20$\tablenotemark{e} & Horizontal Branch Fitting & \cite{ferraro99} \\ 
$13.57\pm0.15$ & Main Sequence Fitting & \cite{reid98} \\ 
$13.32\pm0.03\pm0.04$\tablenotemark{b} & Horizontal Branch Fitting & \cite{kaluzny98} \\
$13.23\pm0.17$ & Baade-Wesselink of RR Lyrae & \cite{storm94} \\
$13.38\pm0.05$ & Horizontal Branch Fitting & \cite{salaris98} \\
$13.52\pm0.08$\tablenotemark{c} & Main Sequence Fitting & \cite{gratton97}\\
$13.28$ & Horizontal Branch Fitting & \cite{hesser87} \\
$13.18$ & Main Sequence Fitting & \cite{hesser87} \\
\enddata
\tablenotetext{a}{Dereddened using E(B~-~V)~$=0.04$ and $A_V/$~E(B~-~V)~$=3.1$.}
\tablenotetext{b}{The first uncertainty is random and the second is systematic.}
\tablenotetext{c}{Derived from the quoted distance modulus uncorrected for binary bias.  A binary correction removes 0.02 mag from the distance modulus.}
\tablenotetext{d}{Using their mean distance derived from two colors.}
\tablenotetext{e}{From their global metallicity derived distance modulus.}
\end{deluxetable*}

The Galactic globular cluster 47 Tucanae (47 Tuc, NGC 104) is considered the metal-rich proto-type for globular cluster work \citep[$\rm{[Fe/H]}=-0.74\pm0.003\pm0.026$, ][]{carretta09}.  
As the closest and least reddened metal-rich globular cluster, 47 Tuc serves as an important science, as well as a calibration target, for the Hubble Space Telescope  (HST).  It is used as a metal-rich anchor when comparing observations of resolved stars in nearby galaxies to stellar evolutionary models \citep[][]{cassisi97,zoccali99,ferraro99,dicecco10} and in examining the star formation history of galaxies in the Local Group  \citep[e.g.][]{monelli10}. 
As another example, the Wide Field Camera~3 (WFC3) Galactic Bulge Treasury program (GO-11664) has targeted 47 Tuc in their resolved stellar population study  \citep{brown09}.  With these observations, ranging from the UV to near-IR, 47 Tuc will serve as an empirical population template, helping to calibrate the reddening-free indices and correcting the transformation of theoretical isochrone libraries into the WFC3 photometric system. 

With relatively well measured values of metallicity and reddening \citep[E(B~-~V)~$=0.04\pm0.02$, ][]{salaris07,harris10}, the distance becomes the crucial input parameter to determine the age of 47 Tuc.  A considerable effort has gone into determining the distance to this cluster.  We list recent work in Table~\ref{tab:distance}.  
There has been quite a variety of methodologies used to determine the distance of this cluster, including main sequence fitting \citep{hesser87,gratton97,reid98,carretta00,grundahl02,percival02,gratton03,bergbusch09}, the tip of the red giant branch \citep{ferraro00,bono08}, horizontal branch fitting \citep{hesser87,kaluzny98,salaris98,ferraro99,salaris07}, RR Lyrae \citep{storm94,bono08}, cluster stellar kinematics \citep{mclaughlin06}, eclipsing binaries \citep{kaluzny07,thompson10}, and white dwarf (WD) cooling sequence fitting \citep{zoccali01}.
This set of measurements spans unreddened distance moduli between $13.02-13.57$ for 47 Tuc which has important implications for the derived age. Examining only the main sequence turnoff luminosity for a cluster at the extremes in distance moduli, this yields a difference in age of $\sim6$ Gyr\footnote{To first order, we find that when examining the turn off magnitude, the age uncertainty is approximatley equal to (10 Gyr/mag)*(uncertainty in the distance modulus).}, altering any inferred history of the formation of the Milky Way that one might derive from 47 Tuc.  

Our goal is to determine the distance to 47 Tuc following the example of \cite{zoccali01} by using the cluster's WD population.  \cite{zoccali01} constructed a WD sequence using local calibrators with measured trigonometric parallaxes and magnitudes to compare with the WD cooling sequence of 47 Tuc, yielding an unreddened distance modulus of (m - M)$_{\rm{o}}=13.15\pm0.14$ (see Table~\ref{tab:distance}).  Here, rather than use local calibrators, we determine the distance by comparing our measured photometry to the hydrogen atmosphere (DA) WD models of \cite{tremblay11} and the helium atmosphere (DB) models of \cite{bergeron11}.  Previous spectroscopic studies have found solely DA WDs in globular clusters within the temperature range of our sample, including the globular clusters M4 \citep{davis09}, NGC 6752 and NGC 6397 \citep{moehler04}, as well as in the open cluster NGC 2099 \citep{kalirai05}.  
While this has not been tested in 47 Tuc to date, we make the assumption that our sample consists of DA WDs, but we also compare our results to that obtained considering DB models and a mixture of DA and DB WDs.

WDs within globular clusters are faint objects in crowded fields, making their identification quite challenging.  However, with carefully planned observations with the HST, combined with thorough data reduction and analysis, it becomes possible to obtain high quality multi-band photometry of WDs in large samples within one cluster system \citep[see the recent observations of WDs in M4 and NGC 6397 from][]{richer02,hansen02,hansen07,richer08}.

\section{Dataset}
\label{sec:dataset}

We obtained HST images of the Galactic globular cluster 47 Tuc in cycle 17 (GO-11677, PI - H. Richer) with the Advanced Camera for Surveys (ACS) and the WFC3.  The primary field of this program was imaged with the ACS/WFC in F606W and F814W filters for 121 orbits. Additional fields were taken with the WFC3 using both the UVIS and IR channels.  Our observational design and reduction are discussed in detail in \cite{kalirai11}, here we briefly review the WFC3 dataset used in the present analysis.  With the WFC3, we obtained 121 orbits covering $>60$ arcmin$^2$ and extending over 250 degrees in azimuthal range of 47 Tuc.  This covers a radial range of $5-13.7$ arcmin ($6.5-17.9$ pc) within the globular cluster.  For 61 orbits, we stared in one field location in the cluster (stare field) obtaining 59 orbits of IR images in F110W and F160W and 2 orbits of UVIS images in F390W and F606W.  We then swept out an arc around the cluster's center (swath fields), using 5 orbits in each of our 12 additional fields obtaining F390W, F606W, F110W and F160W images \cite[for a visualization of our observations, see Figure~4 of][]{kalirai11}.   The observational logs of the swath and stare fields can be found in Tables~2 \&~3 of \cite{kalirai11}.  

To summarize the data analysis discussed in \cite{kalirai11}, we retrieved the WFC3 observations from the MAST archive and re-processed all data into $\_{\rm{flt}}$  images using the latest calibration files.  These images were then supersampled and drizzled together, paying careful attention to image registration, sky offsets, and pixel drop size. PSF-fitted photometry and astrometry was performed on the final image stacks in each filter using DAOPHOT~II and ALLSTAR.  The IR swath data were too undersampled (with a native FWHM$=1.1-1.2$ pixels) for PSF photometry however, so photometry for each star was obtained from a small aperture radius of 3.5 pixels.  The catalogs from all filters, over both the UVIS and IR cameras, were merged together using high order transformations.  

\section{Selecting the Sample}
\label{sec:sample}

\begin{figure}
\plotone{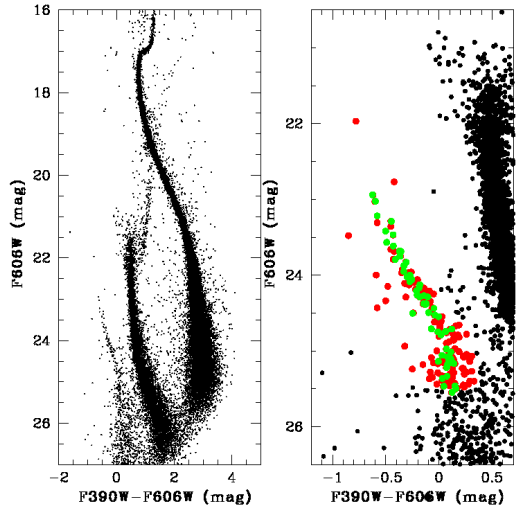}
\caption{{\it Left:} The UVIS CMD of the objects in the stare and swath fields.  The reddest sequence is the main sequence of 47 Tuc and the bluest sequence is the cluster's WD cooling sequence.  The intermediate sequence is the main sequence and turn off region of the SMC in the background.  {\it Right:} A closer examination of the 47 Tuc WD cooling sequence.  Our final selection of 59 WDs are shown as {\it green (light grey) circles} while the remaining 112 candidate WDs are shown as {\it red (dark grey) circles}.  {\it A color version of this figure is available in the electronic version.}}
\label{fig:cmd}
\end{figure}

In the selection of our sample of WDs, our goal was not to have the largest sample possible with the dataset, but instead to have a set of WDs that cleanly lay on the white dwarf cooling sequence and are free of possible photometric contamination.  Therefore, we selected our sample using multiple criteria.   Initially, we isolated the WD cooling sequence in the (F390W-F606W, F606W) CMD shown in Figure~\ref{fig:cmd},
selecting WD candidates down to F606W~$=25.5$. 

Our WD cooling sequence was carefully selected to avoid possible contaminates from the main sequence of the SMC in the background.  From our inital list, there were 171 WD candidates that also had F110W and F160W measurements.  From these 171 candidates, we selected objects that were consistent with being stellar {\it and} passed a visual inspection of the images.  

\begin{figure}
\plotone{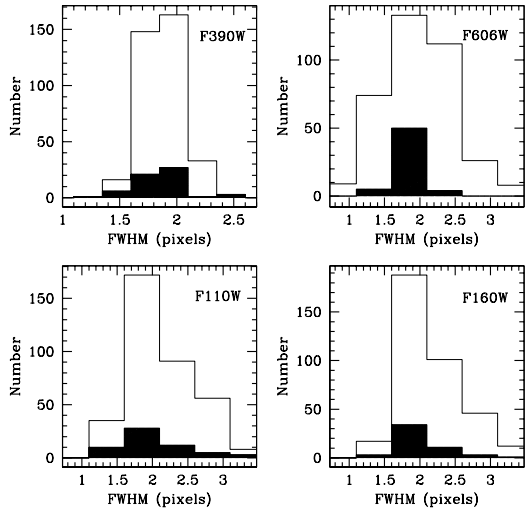} 
\caption{The measured FWHMs for main sequence stars between F606W~$=19-19.5$ magnitudes in all fields are shown as the open histogram for the filters ({\it top left}) F390W, ({\it top right}) F606W, ({\it bottom left}) F110W, and ({\it bottom right}) F160W.  The solid histograms are the measured FWHMs for the 59 selected WDs.}
\label{fig:fwhm}
\end{figure}

To classify an object as stellar, {\it full width half maxima} (FWHMs) were measured in all 4 images and compared to the mean stellar FWHM in these same images. To determine the mean stellar FWHM in each image, we measured the FWHM of 20-30 stars that lay on the main sequence between F606W~$=19-19.5$ in the (F390W-F606W, F606W) CMD.  We selected the same 20-30 stars for the 4 filters in each field.
We removed candidate objects from our list of 171 candidates that differed in their measured FWHM by more than 1 standard deviation from the mean stellar value in the F160W filter.  We chose the F160W filter because the resolution of the images are the lowest in our sample and the WDs are thus more likely to be contaminated by nearby neighbors.  
We are also interested in searching for IR excess in the WD photometry.  This may be indicative of a debris disk or a low mass companion (see Section~\ref{sec:irexcess}).  While most of the IR excess, if present, will be evident at wavelengths longer than F160W, this program could be able to see hints of the excess in the F160W band which may be followed up in future observations.  As such,  we are particularly critical of the F160W measurements.  The distributions of stellar FWHMs and the final selected WDs are shown in Figure~\ref{fig:fwhm}.  
\begin{figure}
\plotone{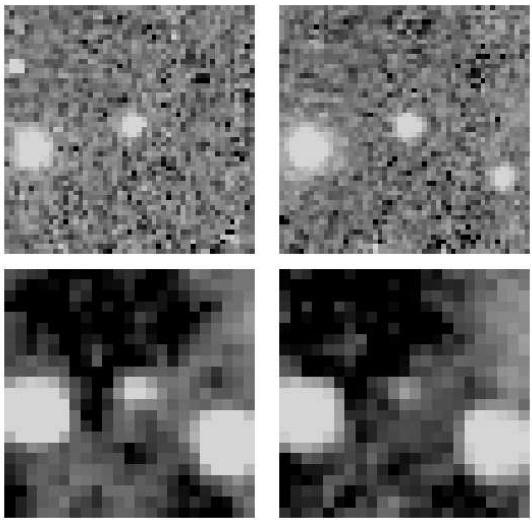}
\caption{A four panel image of WD 1. Within the four panels, the {\it upper left} is F390W, {\it upper right} is F606W, {\it lower left} is F110W, and {\it lower right} is F160W.  The WD is centered in the 2\arcsec~x~2\arcsec\ images and its photometry is listed in Table~\ref{tab:wds}.}
\label{fig:images}
\end{figure}

Independent of our size rejection analysis, we also visually examined the F390W, F606W, F110W, and F160W images for the 171 WD candidates to search for objects that appeared elliptical or whose photometry may have been contaminated by bright neighboring stars, diffraction spikes, remaining cosmic rays, or those found in extremely crowded regions with varying background.  If the candidate object was contaminated in any of the four filters, it was rejected from our list.  

We found in these two search methods that all of our visually rejected objects were also rejected using our FWHM technique.  For a few objects, we were unable to measure a reliable FWHM in F160W, so we accepted these objects into our final list if their FWHMs were stellar in the three additional bands and they did not appear contaminated on the images.  Our final selection consists of 59 WDs listed in Table~\ref{tab:wds} with their field location, R.A. and Decl. (in J2000 epoch), their F390W, F606W, F110W, and F160W photometric measurements and uncertainties in VEGAMAGS, and the effective temperatures that we fit in this study.  In Figure~\ref{fig:images}, we show the F390W, F606W, F110W, and F160W images for our sample WDs.  Two of these WDs were found near the edge of their F390W and F606W images.  For these two filters, there were only two exposures combined to produce the final image and thus some cosmic rays were not removed due to the offset between the two exposures. This is evident in Fig.~\ref{fig:images} for WDs 23 and 52 but does not affect our photometry in these bands obtained with PSF fitting.

\section{Spectral Energy Distributions}
\label{sec:seds}
\subsection{Model Conversion to our Photometric System}
\label{sec:modconvert}

In order to generate the spectral energy distributions of WDs in 47 Tuc, we have chosen to use the \cite{tremblay11} spectral model grids of pure hydrogen atmosphere WDs.  These model grids define spectra for objects with surface gravities (log(g)) ranging from $6-10$ in steps of 0.5 and effective temperatures (T$_{\rm{{eff}}}$) from $6 000-120 000$~K.  The spectral grid is defined in steps of $500$~K between $6 000-17 000$~K, steps of $5 000$~K between $20 000-90 000$~K, and steps of $10 000$~K between $90 000-120 000$~K.  We have converted the model grids to our observed Space Telescope magnitudes (STMAG), similar to the procedure of \cite{bergeron97}, to obtain a magnitude for each of our observed filters, $M_{\rm{STMAG}}$.  This enables us to fit the models to our observations via Equation~\ref{eqn:mag},
\begin{equation}
\label{eqn:mag}
M_{\mathrm{STMAG}}=-2.5\log_{10}\!\left(\frac{\frac{R^2}{d^2} \int_0^\infty \! \lambda F_{\mathrm{mod}} E_{\lambda} S_{\lambda} \, \mathrm{d}\lambda}{\int_0^\infty \! \lambda  F_{0} S_{\lambda} \, \mathrm{d}\lambda}\right)
\end{equation}
where $F_{\mathrm{mod}}$ is the flux per unit wavelength for the spectral model integrated over the solid angle, $S_{\lambda}$ is the total system throughput for the filter\footnote{The STMAG zeropoints and throughput values for each filter have been obtained from the {\it Space Telescope Science Institute} website (http://www.stsci.edu/hst/wfc3/phot$\_$zp$\_$lbn, http://www.stsci.edu/hst/wfc3/ins$\_$performance/throughputs).}, and $E_{\lambda}$ is the standard interstellar extinction curve \citep{fitzpatrick99} with $A_V/$~E(B~-~V)~$=3.1$, using some input E(B~-~V).  The STMAG zeropoints are defined so that a flat spectrum of 3.63\e{-9} ergs cm$^{-2}$ s$^{-1}$ $\rm{\AA}^{-1}$ ($F_0$) through any filter yields a magnitude of zero. 

The inclusion of the $\lambda$ terms in Equation~\ref{eqn:mag} is required because of how the flux is measured on a CCD.  The CCD does not know the individual energy of each photon that it receives, only whether or not it has received a photon.  For any given energy density, the number of photons received will increase linearly with wavelength.  For a relatively flat spectrum, fluxes measured on a CCD through some filter will therefore be more affected by the energy flux at longer wavelengths.  This is because this region of the spectrum will necessarily have a higher number of photons per unit energy as the energy of a photon is inversely proportional to its wavelength ($\rm{E_{photon}}={ch}/{\lambda}$).

\begin{figure*}
\plotone{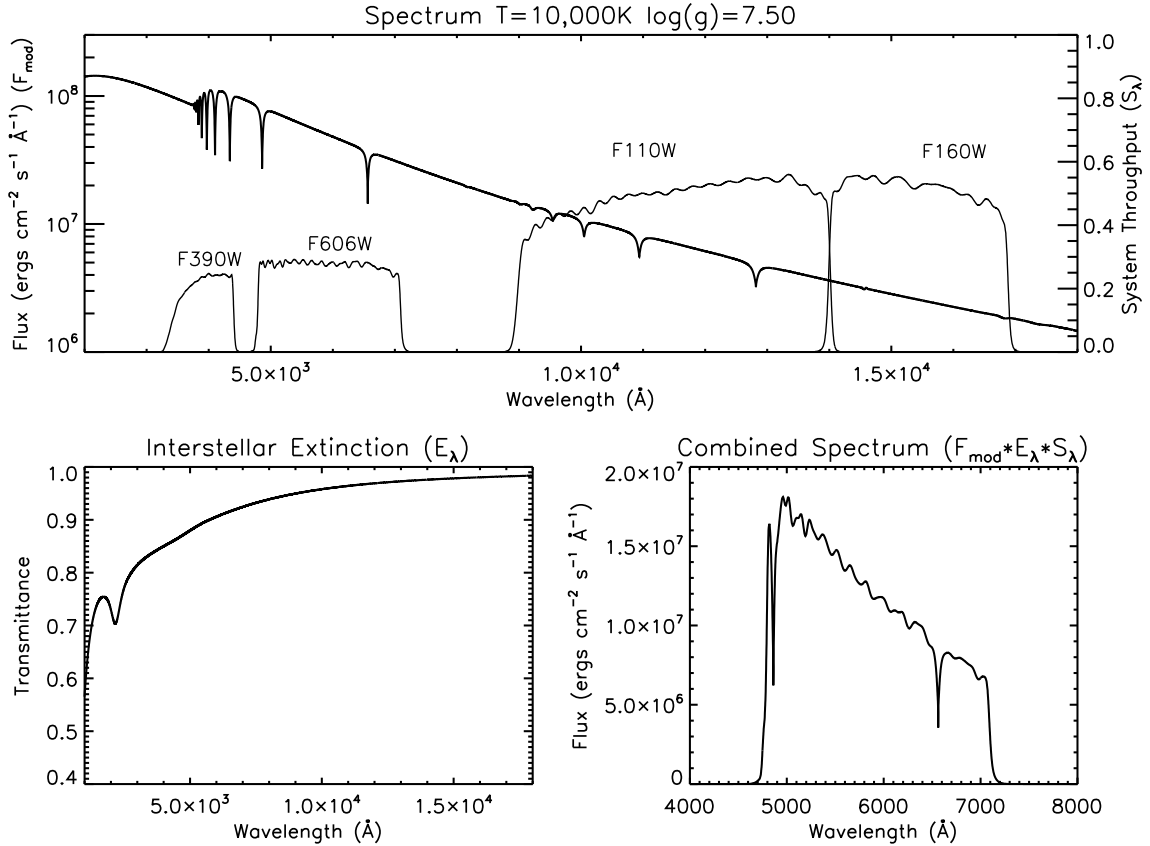} 
\caption{An illustration of Equation~\ref{eqn:mag2} for a sample model spectrum converted to an F606W magnitude.  {\it Upper panel}: An example model spectrum from \cite{tremblay11} ($F_{mod}$) for a DA WD with an effective temperature of 10 000~K and surface gravity of 7.50. Overlaid are the four filter throughputs $S_{\lambda}$ (F390W, F606W, F110W, and F160W).  {\it Lower left}:  The interstellar extinction. {\it Lower right}: The combined spectum. }
\label{fig:spectomag}
\end{figure*}

The factor of $\rm{(R/d)}^{2}$ in Equation~\ref{eqn:mag} scales the model flux, the flux per unit wavelength at the surface of the object, to the observed flux for a star of radius R, at a distance d.  Equation~\ref{eqn:mag} can then be rewritten as,
\begin{eqnarray}
\label{eqn:mag2}
M_{\mathrm{STMAG}}=-2.5\log_{10}\!\left(\frac{\int_0^\infty \! \lambda F_{\mathrm{mod}} E_{\lambda} S_{\lambda} \, \mathrm{d}\lambda}{\int_0^\infty \! \lambda F_{0} S_{\lambda} \, \mathrm{d}\lambda}\right) \\
+5\log_{10}\!\left(\frac{d}{R}\right) \nonumber
\end{eqnarray}
Equation~\ref{eqn:mag2}, which is illustrated in Figure~\ref{fig:spectomag}, 
shows that the model magnitude in any given filter is a function of only log(g), T$_{\rm{{eff}}}$, reddening, and d$/$R.  We apply this method to convert the model grid magnitudes to our four filters (F390W, F606W, F110W, and F160W).    

\subsection{Fitting the Model SEDs to the WD Photometry}
\label{sec:wdphot}

\begin{figure}
\plotone{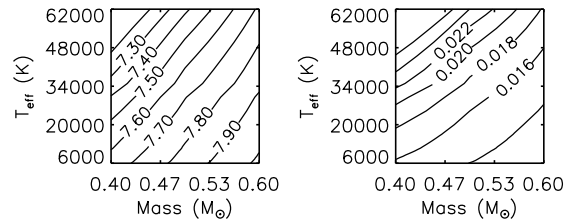}
\caption{{\it Left}: Contours of constant surface gravity (log(g)) in mass-effective temperature (T$_{\rm{eff}}$) space and {\it Right}: Contours of constant radius, measured in units of solar radii, in mass-T$_{\rm{eff}}$ space from \cite{tremblay11}.}
\label{fig:massconts}
\end{figure}

Using WD cooling models from \cite{fontaine01}, we can derive a relationship between surface gravity and effective temperature for a given mass (see Figure~\ref{fig:massconts}).  
Additionally, for any given mass and log(g), we can calculate the radius from Equation~\ref{eqn:radius}.  
\begin{equation}
\label{eqn:radius}
R=\sqrt{\frac{GM}{g}}.
\end{equation}
Putting the mass dependent T$_{\rm{eff}}$ - log(g) relation together with Equation~\ref{eqn:radius} means that, of the four parameters, T$_{\rm{eff}}$, log(g), mass, and radius, the combination of T$_{\rm{eff}}$ and any of the other three allows one to calculate the remaining two parameters.  Of these four, we have chosen to parametrize our grid over T$_{\rm{eff}}$ and mass, since previous work allows us to put a reasonable prior on the mass of our objects.
This reduces the six parameters that control each WD spectrum (effective temperature, mass, distance, reddening, surface gravity, and the radius of the WD) to four parameters that need to be fit in our procedure.  We are therefore able to generate a model SED in these four filters given a mass, effective temperature, reddening, and a distance.  The first three parameters control the shape of the SED, while the distance controls the overall scaling. In
order to determine temperatures for our WDs and to display both model and WD SEDs, we have first held mass and reddening as constants in our fitting procedure.  We have assumed a mass of $0.53$ M$_{\odot}$.  This follows from the the prediction of \cite{renzini88} and \cite{renzini96} that the WDs forming today in globular clusters should have a mass of $0.53 \pm 0.02$ M$_{\odot}$.  Further support of our mass selection comes from recent direct spectroscopic measurements from \cite{kalirai09} of $0.53 \pm 0.01$ M$_{\odot}$ for the mass of WDs in M4 and $0.53 \pm 0.03$  M$_{\odot}$ for the mass of WDs in NGC 6752 \citep{moehler04}.  

\begin{figure}
\plotone{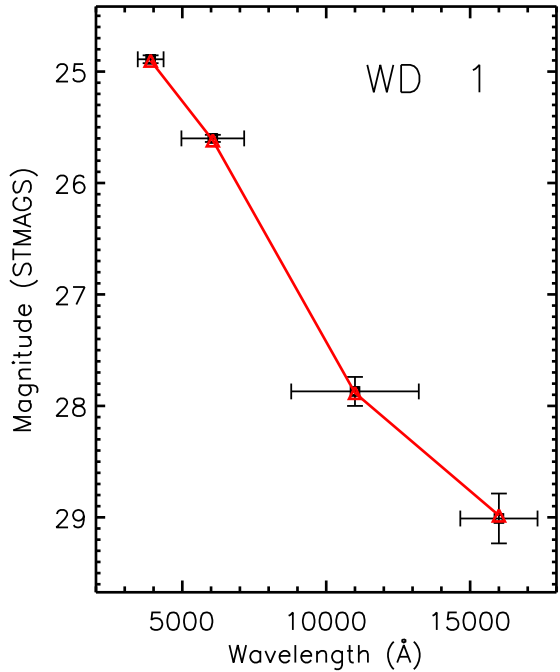}
\caption{The SED for WD 1.  The {\it filled squares} are our photometric measurements in F390W, F606W, F110W, and F160W with uncertainties.  The wavelength range for each filter is also shown as the equivalent width of the filter divided by the maximum throughput, as defined by the WFC3 instrument handbook.  The {\it red triangles}, connected by the {\it red line} is the best fit model in STMAG magnitudes. {\it A color version of this figure is available in the electronic version.}}
\label{fig:sed_prob}
\end{figure}

We also assume a reddening of $0.04$ and then fit the distance modulus and the effective temperature to each individual object.  We ignore the individual distance moduli here and include the fitted effective temperatures in Table~\ref{tab:wds} for our 59 WDs. Figure~\ref{fig:sed_prob} shows the SEDs for our selected WD sample.   

In Section~\ref{sec:distance}, we determine the distance to 47 Tuc using Equation~\ref{eqn:mag2} to fit to our WD photometry, however we do so using a maximum likelihood function and marginalise over various parameters.

\section{Distance to 47 Tuc}
\label{sec:distance}

To determine the distance to 47 Tuc, we will use white dwarf spectral models in order to predict an expected absolute magnitude for our sample WDs in our 4 filters, similarly to our technique in Section~\ref{sec:seds}.  However in this section, we fit the distance modulus to 47 Tuc by comparing the expected absolute magnitude to the measured apparent magnitudes of the WDs using a maximum likelihood analysis that uses all 59 WDs to determine one final distance to the cluster. Unlike in Section~\ref{sec:seds}, we do not assume a single value for reddening and WD mass in the determination below, rather we marginalise over each of these parameters with the inclusion of Gaussian priors.
\vspace{1cm}
\subsection{Maximum Likelihood Function}
\label{sec:likelihood}

Our procedure begins by calculating the likelihood function of our 59 WDs, termed $L_j$ where $j$ runs from $1-59$, over our four parameters of T$_{\rm{eff}}$, reddening, mass, and distance which is simply converted to a true distance modulus, (m - M)$_{\rm{o}}$. This is shown in Equation~\ref{eqn:likelihood1}
\begin{eqnarray}
\label{eqn:likelihood1}
L_{j}(data\mid(\mathrm{T_{\mathrm{eff}},(m-M)_{o},mass,E(B-V)}))= \\
\prod_{i=1}^4\mathrm{exp}\!\left[-\frac{(data_{i}-mod_{i})^{2}}{2\sigma_{i}^{2}}\right]. \nonumber
\end{eqnarray}
The photometric data for each WD (termed $data_i$) are the measured magnitudes in our 4 filters, $mod_i$ are the magnitudes calculated from the model described in Equation~\ref{eqn:mag2} from Section~\ref{sec:seds}, and $\sigma_i$ are the uncertainties for each point.

After a likelihood for each white dwarf is calculated using Equation~\ref{eqn:likelihood1}, we use Bayes' Theorem as well as a Gaussian prior on the mass with a mean of 0.53 M$_{\odot}$ and a standard deviation of 0.02 M$_{\odot}$, and a uniform prior on the temperature, to calculate the likelihood of the model parameters (m - M)$_{\rm{o}}$ and E(B-V) for each individual object, given the data.  This step is described in Equation~\ref{eqn:likelihood2}

\begin{eqnarray}
\label{eqn:likelihood2}
 L_{j}((\mathrm{(m-M)_{o},E(B-V)})\mid data) \propto \\
 \int_{0}^{\infty} \int_{0}^{\infty} L_{j}(data\mid(\mathrm{T_{\mathrm{eff}},(m - M)_{o},mass,E(B-V)})) \: \nonumber \\
\mathrm{*exp}\!\left[-\frac{(0.53-mass)^{2}}{2(0.02)^{2}}\right] \: \mathrm{dmass} \: \mathrm{d}\rm{T_{eff}}. \nonumber
\end{eqnarray}
 
\begin{figure}[H]
\plotone{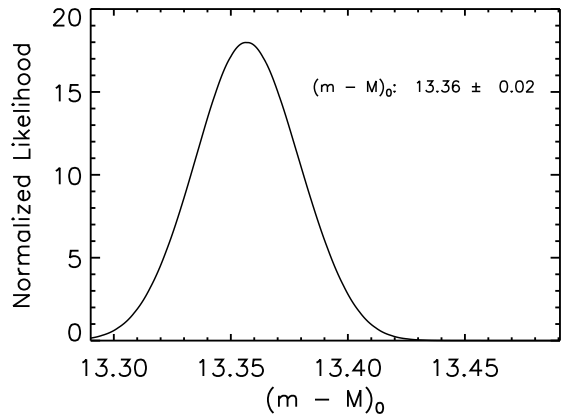}
\caption{The combined distribution function of distance modulus for the sample of 59 WDs.  The peak of the distribution at $13.36$ is the unreddened distance modulus value we quote for 47 Tuc along with its statistical uncertainty of $0.02$.}
\label{fig:dm}
\end{figure}

We then combine the distance modulus-reddening likelihood of all 59 WDs described in Equation~\ref{eqn:likelihood3}
\begin{eqnarray}
\label{eqn:likelihood3}
L((\mathrm{(m-M)_{o},E(B-V)})\mid data)= \\
\prod_{j=1}^{59} L_{j}((\mathrm{(m-M)_{o},E(B-V)})\mid data) \nonumber
\end{eqnarray}
and marginalise over the reddening parameter with a Gaussian prior with a mean of $0.04$ and a $\sigma=0.02$.  From this, we obtain the distribution function for the cluster as a whole, described in Equation~\ref{eqn:likelihood4} and shown in Figure~\ref{fig:dm}. 
Our best fit unreddened distance modulus using these priors on mass and reddening is $13.36\pm0.02$. 
\begin{eqnarray}
\label{eqn:likelihood4}
 L((\mathrm{(m-M)_{o}})\mid data)= \\
\int_{0}^{\infty} L(\mathrm{(m-M)_{o},E(B-V)})\mid data) \: \nonumber \\
\mathrm{*exp}\!\left[-\frac{(0.04-\mathrm{E(B-V)})^{2}}{2(0.02)^{2}}\right] \: \mathrm{d E(B-V)} \nonumber
\end{eqnarray}

Additionally, we consider how our distance estimate depends on our assumptions about the mass and reddening.  To do this, we can consider our original four parameter likelihood grid for each of our 59 white dwarfs.  After marginalising out the effective temperature for each WD individually and combining the likelihood of all 59 WDs in Eqn.~\ref{eqn:likelihood3}, we are left with the likelihood values on a grid over reddening, mass, and distance modulus.  For each combination of reddening and mass on the grid, there is a corresponding one-dimensional distance modulus likelihood distribution.  The most likely distance modulus can then be found for each mass and reddening pair.  The result is a surface of most likely distance modulus values that is well approximately by a plane, described as 
\begin{eqnarray}
\label{eqn:dmeqn}
\rm{(m-M)_{o}} = 14.634 - (2.573 *\rm{mass}) \\
 + (2.552 *\rm{E(B-V)}) \nonumber
\end{eqnarray}
where the mass is in solar masses.  This approximation to our likelihood grid is accurate to 0.003~mag and is valid only within the ranges of mass ($0.50-0.56$ M$_{\odot}$) and reddening ($0-0.08$).

\subsection{Fitting the SEDs with DB Atmosphere Models}
\label{sec:dbmodels}

Here we test the assumption that the WDs in our sample are DA atmosphere WDs by fitting our sample of 59 WDs instead with DB atmosphere models \citep{bergeron11}.  The SEDs fit with the DB models return temperatures that are on average $2200$~K cooler than the fits with the DA models.  These cooler temperatures result in objects that are fainter and thus our true distance modulus is closer at a value of $12.90$.  If we compare the likelihoods of these fits, 53 of our 59 WDs have a higher likelihood value for our best fit DA model, with 48 being better by a factor of 10 or more and 22 being better by a factor of $10^4$ or more.  The 6 remaining WDs that were fit better by a DB model have an average likelihood that is better by a factor of 2.  Additionally, the standard deviation of the differences between the best fit model and all 59 WDs is $0.11$~mag for the DA models and $0.18$~mag for the DB models.  

In the field population of WDs, \cite{tremblay08} found the ratio of DB to DA to be approximately $25\%$ in the temperature range of our WD sample.  We test the effect on the determined distance modulus if $25\%$ of our sample were instead DB WDs, by randomly fitting $25\%$ of our WDs by DB models and the rest with DA models.  Under these conditions, the distance modulus is $13.32$ with a random uncertainty of $0.06$, easily encompassing our distance modulus assuming all DA WDs. 

\subsection{Error Analysis}
\label{sec:errors}

\begin{figure}
\plotone{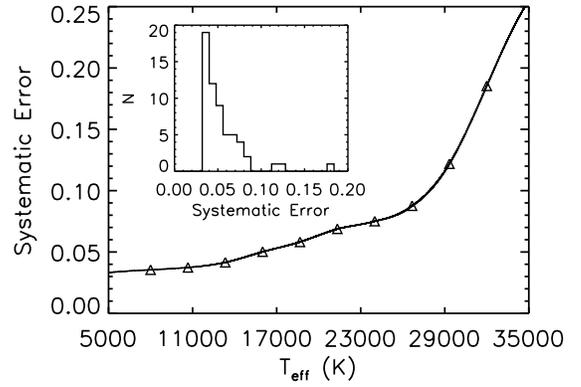}
\caption{The systematic uncertainties of the distance modulus as a function of effective temperature (T$_{\rm{eff}}$) for 10 synthetic WDs ({\it triangles}) fit with a cubic spline function.  {\it Inset}: The distribution of systematic errors for the 59 WDs in our sample determined from the spline fit.  Our quoted systematic uncertainty for the distance modulus of 47 Tuc is the mean of the sample of 59 WDs.}
\label{fig:errors}
\end{figure}

The statistical error of $0.02$ that is quoted from our maximum likelihood analysis in Section~\ref{sec:likelihood} are $1\sigma$ upper and lower errors for the distance modulus calculated directly from the summed likelihood distribution in Fig.~\ref{fig:dm}.
The systematic errors were determined entirely from the uncertainties in our photometric calibration taken from \cite{kalirai11}.  We generate 10 synthetic WD SEDs from the models of \cite{tremblay11} evenly spaced between $8000-32000$ K, 
spanning the range of effective temperature of our WD sample. We performed the same likelihood analysis as in Section~\ref{sec:likelihood} using the uncertainties in the photometric calibration as the errors on each magnitude.  For each of the 10 WDs, we extract a distance modulus from their likelihood distributions along with the width of the distribution as the error.  The widths of the likelihood distributions must be due solely to our photometric calibration uncertainties, as these synthetic objects come directly from the model grids.  Figure~\ref{fig:errors} shows the systematic errors as a function of effective temperature for the synthetic WDs fit with a cubic spline function. For each of our 59 WDs, we used the effective temperature to obtain the systematic error on its determined distance modulus from the spline fit in Fig.~\ref{fig:errors}.For the final calibration error on the distance modulus of 47 Tuc, we quote the mean systematic error of the 59 WDs.    

\begin{figure}
\plotone{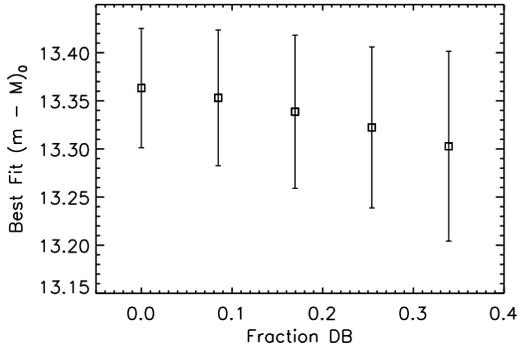}
\caption{The best fit distance modulus shown with varying fractions of DB/DA WDs in our sample. The uncertainties shown are the random and systemic uncertainties added in quadrature.}
\label{fig:dbfraction}
\end{figure}

Using this technique, we have determined the true distance modulus to 47 Tuc to be $13.36 \pm0.02$(random)$ \pm 0.06$(systematic) corresponding to a distance of $4.70\pm0.04$(random)$\pm0.13$(systematic) kpc. An unknown fraction of DB objects in our sample will systematically change this result, giving us a smaller distance modulus, and wider error bars as the fraction increases.  The closer distance results from the fact that DB white dwarfs of roughly the same SED shape are brighter than DAs, meaning we would  have to fit them closer in order to match the observed magnitudes.  The wider error bars result from the fact that DB models almost universally fit our objects worse than DA models.  Additionally, we don't know which objects are DBs, and so this causes a systematic error that needs to be addressed by iterating our fitting routine while randomly assigning some fraction of our objects to be fit as DBs rather than DAs each time.  The fit distance modulus and errors both change roughly linearly with DB fraction, as shown in Figure~\ref{fig:dbfraction}.  
At $0\%$ DBs we have the quoted fit values, at $25\%$ DBs we have $13.32\pm0.06\pm0.06$. 

\begin{figure}
\plotone{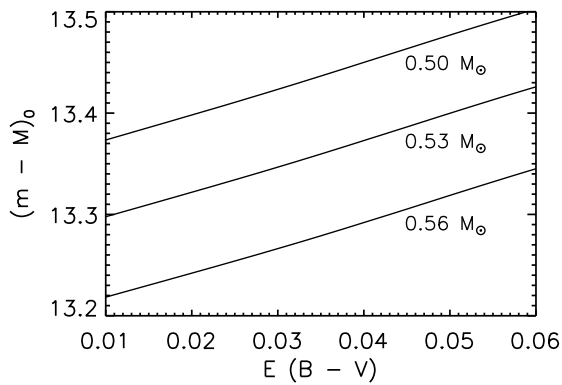}
\caption{The distance modulus determined from our maximum likelihood technique shown as a function of reddening for various slices in WD mass space.}
\label{fig:variations}
\end{figure}

We have tried a number of variations in order to assess how the distance modulus changes by varying some of our input assumptions.  The white dwarfs forming in globular clusters today should have masses between $0.51-0.55$~M$_{\odot}$ \citep{renzini88,renzini96}.  As discussed in Section~\ref{sec:likelihood}, we have used a Gaussian prior on the mass with a mean of 0.53 M$_{\odot}$ and a $\sigma=0.02$ in our distance determination.  Based on Eqn~\ref{eqn:dmeqn}, if we instead assume a WD mass of 0.50~M$_{\odot}$ and 0.56~M$_{\odot}$, our unreddened distance moduli becomes $13.45$ and $13.30$, respectively.  The reddening towards 47 Tuc has estimates ranging between E(B~-~V)~$=0.024$ \citep{gratton03} and E(B~-~V)~$=0.055$ \citep{gratton97} in the literature.  
If we fix the reddening values to E(B~-~V)~$=0.024$ and E(B~-~V)~$=0.055$, we obtain unreddened distance moduli of $13.33$ and $13.41$, respectively.
Despite various changes to our initial assumptions, these are all consistent with our distance modulus of $13.36$ given the random uncertainties of $0.02$ on each distance and the systematic uncertainty of $0.06$ quoted above.  We show in Figure~\ref{fig:variations} 
how the distance modulus changes as a function of reddening taken at different WD masses, all spanning the ranges in the literature.  Again, we find all reasonable assumptions of the initial parameters to be consistent within our quoted uncertainties.
\begin{figure}
\plotone{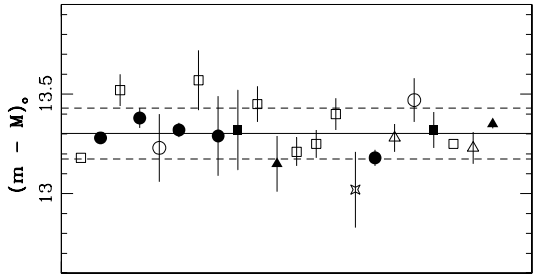}
\caption{The previously determined unreddened distance moduli of 47 Tuc including the value from this study, plotted with their statistical uncertainties.  The distances plotted from right to left are listed in Table~\ref{tab:distance} from top to bottom. Our new measurement is the final point on the right.  The different symbols represent the various methods used in the distance determination:  main sequence fitting ({\it open squares}), horizontal branch fitting ({\it filled circles}), RR Lyrae ({\it open circles}), tip of the red giant branch ({\it closed squares}), methods using the white dwarfs ({\it filled triangle}), cluster kinematics ({\it cross}), and eclipsing binaries ({\it open triangles}). The {\it solid line} is the mean of the 22 distance moduli measurements ($<$~(m~-~M)~$_{\rm{o}}>=13.30$) and the {\it dashed lines} are one standard deviation ($\sigma=0.13$).}
\label{fig:distance}
\end{figure}

As a summary, we show the previous distance determinations to 47 Tuc in Table~\ref{tab:distance} including the distance determined here in Figure~\ref{fig:distance}. These values have been dereddened, where appropriate, using E(B~-~V)~$=0.04$ and $A_V/$~E(B~-~V)~$=3.1$.  We find the mean of these measurements to be $<$~(m~-~M)~$_{\rm{o}}>=13.30$ with a standard error of the mean of $0.03$ and a standard deviation of $0.13$.

\section{Discussion}
\label{sec:conclusions}

\subsection{A Search for Infrared Excess}
\label{sec:irexcess}

Our dataset also provides us the opportunity to search for evidence of extrasolar planetary systems.  A previous search for planets in the core of 47 Tuc was performed by \cite{gilliland00} using a time-series analysis over $8.3$ days searching for close in Jupiter-sized planets around $\sim34 000$ main sequence stars.  They concluded that no light curve in their sample strongly suggested the presence of a planet, indicating that the planet frequency in 47 Tuc may be an order of magnitude less than in our solar neighborhood.  This may be expected in the dense stellar environment of the core of 47 Tuc which could lead to orbit disruption and/or hinder planet survival.  In addition, the planet-metallicity correlation indicates that the frequency of planets increases with the host star's metallicity \citep[e.g.][]{fischer05}, making it even less likely to find a planet around a star with subsolar metallicities in 47 Tuc. 
\begin{figure}[H]
\plotone{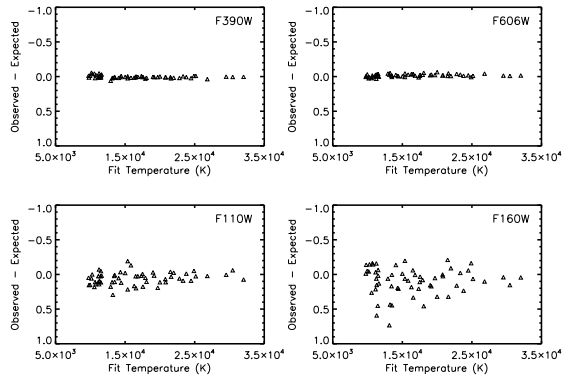}
\caption{The observed minus the expected model magnitudes as a function of effective temperature (T$_{\rm{eff}}$) for our WD sample in the four filters F390W ({\it upper left}), F606W ({\it upper right}), F110W ({\it lower left}), and F160W ({\it lower right}). }
\label{fig:obs_exp}
\end{figure}

Our dataset enables us to instead search for evidence of extrasolar planetary systems with the SEDs of our selected WDs.  WDs have essentially metal-free atmospheres \citep{paquette86,zuckerman03} and low luminosities, allowing for  the detection of cool low mass counterparts, such as debris or circumstellar disks, as well as planets or brown dwarfs \citep{probst83,zuckerman87a,farihi05,burleigh06}. One form of evidence for the existence of these features is the presence of IR excess \citep[][among others]{zuckerman87b,graham90,kilic06,vonhippel07,farihi08}.   
The likelihood of a WD having an IR excess is correlated with its T$_{\rm{eff}}$ and/or presence of photospheric metals \citep{farihi09}, the most well studied of which is calcium.  The pollution of hydrogen-rich photospheres of WDs with metals could be caused by circumstellar accretion.  If a planetesimal or asteroid is tidally disrupted, a circumstellar dust disk could arise \citep{jura03,jura08} and likely pollute the WD's atmosphere with detectable metals. 

With no prior information about the calcium abundance of the WDs in our study, we examine their T$_{\rm{eff}}$.  Of our 59 WDs, all but 3 have fitted T$_{\rm{eff}}>10 000$~K, with the remaining three having T$_{\rm{eff}}=9684\pm1041$~K, T$_{\rm{eff}}=9804\pm921$~K, and $9924\pm841$~K.  We can estimate how many cases of IR excess we expect to see in our sample using upper and lower bounds provided by various studies in the literature.  
Recent work by \cite{zuckerman10} show that $\sim30\%$ of the photospheres of white dwarfs with T$_{\rm{eff}}$ between $13 500-19 500$~K show evidence for being polluted with metals. In the same temperature range, approximately $20\%$ of DA WDs with metal bearing photospheres also show evidence for warm circumstellar material \citep{kilic08}.  This indicates a fraction of dusty WDs to be $6\%$, which we take as an upper limit.  A lower limit to the fraction of warm DA WDs with dusty disks is $\sim1\%$ \citep{farihi09}, obtained from recent large WD surveys \citep{liebert05,koester05,farihi05,hoard07,mccook99}.  From our 171 (59) WDs, we therefore expect between $\sim2-10$ ($\sim0-4$) WDs to have IR excess.  However, our reddest photometric measurement is F160W, a wide-H band filter with a central wavelength of $1536.9$~nm.  This is just where the spectral energy distribution of a typical warm WD would begin to show evidence for an IR excess given a companion or disk with T$_{\rm{eff}}\sim800$~K.  In addition, we must also consider our photometric uncertainties. It thus seems unlikely that we would find clear evidence for an IR excess in our best sample of WDs.  However, if we do find indication for IR-excess in the H band, it could be confirmed with additional observations in redder filters.

\begin{figure}
\plotone{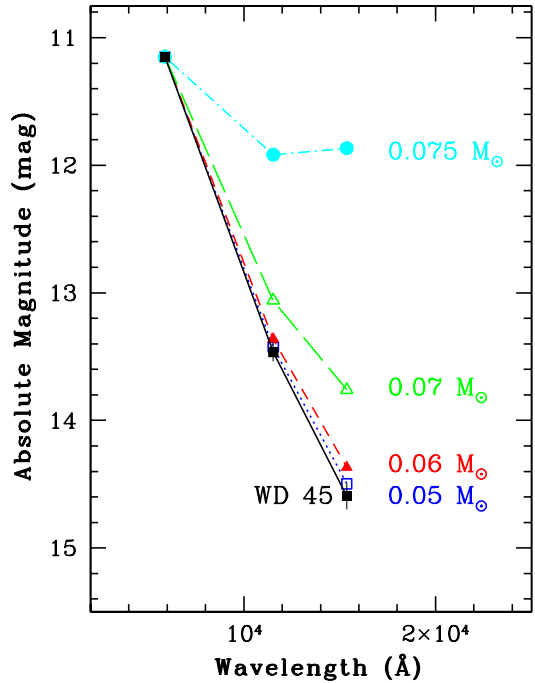}
\caption{The observed F606W, F110W, and F160W magnitudes for a typical WD in our sample ({\it solid squares}).  Also shown are the added contribution to these magnitudes from a 10 Gyr dust free cool brown dwarf companion with masses of $0.05$ M$_{\odot}$ ($\sim50$ M$_{J}$) as {\it open blue squares}, $0.06$ M$_{\odot}$ ($\sim60$ M$_{J}$) as {\it solid red triangles}, $0.07$ M$_{\odot}$ ($\sim70$ M$_{J}$) as {\it open green triangles}, and $0.075$ M$_{\odot}$ ($\sim75$ M$_{J}$) as {\it solid cyan circles} from the \cite{baraffe03} models. Here, we have converted the model magnitudes to that of 2MASS using the transformations in \cite{carpenter01}. In order for us to detect a cool brown dwarf companion from our sample, the companion must be at least $0.06$ M$_{\odot}$, which is an excess of $0.23$ magnitudes in the F160W filter. {\it A color version of this figure is available in the electronic version.}}
\label{fig:sed_companion}
\end{figure}

We have searched the SEDs in Figure~\ref{fig:sed_prob} of all our WD candidates for evidence of an IR excess, focusing on the examination of the F160W band.  Any evidence that we found for an enhanced F160W measurement appeared to be due to contamination by neighboring stars, diffraction spikes, or cosmic rays.  In addition, for each filter, we plot the observed magnitude minus the expected model magnitude as a function of effective temperature for each of our 59 WDs, shown in Figure~\ref{fig:obs_exp}. 
The few outlying points in the F160W filter at lower temperature are objects that are too faint compared to the models, driven by large photometric uncertainties.  If IR excess is present, we would expect to see points with negative values on the y-axis.  Figure~\ref{fig:obs_exp} indicates that in the F160W band, our WDs are at most in excess of 0.2 magnitudes compared to the model.  We question whether this could indicate a cool brown dwarf companion or debris disk.  Using the evolutionary models for cool brown dwarfs and exosolar giant planets of \cite{baraffe03}, we combine a typical WD in our sample with the contribution from a 10 Gyr dust free cool brown dwarf. Here, we convert the model IR data from the CIT system to that of 2MASS using the transformations of \cite{carpenter01}, as 2MASS is a nearly identical photometric system to that in WFC3 and provides the closest comparison possible.   In Figure~\ref{fig:sed_companion}, we show that for masses of a companion below $0.06$ M$_{\odot}$ ($\sim60$ M$_{J}$), we would be unable to detect it. However, a cool brown dwarf with $0.06$ M$_{\odot}$, may be detectable, as this leads to a difference in magnitude in the F160W band of $>0.23$ magnitudes. 
 Examining Figure~\ref{fig:obs_exp}, it is clear that none of our 59 WDs have F160W in excess of 0.2 from the models.
We conclude that we do not find any case in our sample that shows strong evidence for a genuine IR excess.   Our search for planets in 47 Tuc continues, however, with this cycle 17 dataset.  We have time series data of two fields outside the core of 47 Tuc, one is the stare field described in Section~\ref{sec:dataset} consisting of F110W and F160W observations over $\sim20$ days with $10000$ stars, and the other field has ACS images in F606W and F814W filters spanning $\sim9$ months with $20000$ stars.  Our analysis, currently underway, will be able to put a constraint on the binary frequency within 47 Tuc and search for exoplanets by analysis of the stellar light curves \citep{woodley12}. Not only will we examine the light curve of main sequence stars down to $\sim0.2$ M$_{\odot}$, but we will also be able to search for planets around the brighter WD population.  A small radius ratio between the WDs and giant gas planets will significantly aid in our planetary search.

\subsection{Summary}
\label{sec:summary}

With this new HST dataset of the Galactic globular cluster 47 Tuc, we have examined the SEDs of a sample of 59 WDs that are devoid of photometric contamination by neighboring stars or image defects.   From this sample of SEDs, we have applied a novel approach to simultaneously estimate the distance of 47 Tuc from the comparison of theoretical model WD atmospheres \citep{tremblay11}.  We obtain a true distance modulus of 
$13.36\pm0.02$(random)$\pm0.06$(systematic) corresponding to a distance of $4.70\pm0.04$(random)$\pm0.13$(systematic)~kpc.  

\acknowledgements We thank our referee who provided a very careful, comprehensive, and thoughtful report. We extend our thanks to Dr. Zoccali for providing us with local white dwarf photometry.  Support for the program GO-11677 was provided by NASA through a grant from the Space Telescope Science Institute, which is operated by the Association of Universities for Research in Astronomy, Inc., under NASA contract NAS~5-26555.  This work is supported in part by the NSERC Canada and by the Fund FQRNT (Qu\'ebec)


\clearpage
\LongTables
\begin{deluxetable}{rlrrrrrrrrrrrr}
\tablecolumns{14}
\tabletypesize{\small}
\tablecaption{Data on the Selected White Dwarfs\label{tab:wds}}
\tablewidth{0pt}
\tablehead{
\colhead{ID}&\colhead{Field}&\colhead{$R.A.$}&\colhead{$Decl.$} &\colhead{F390W} &\colhead{$\sigma_{\rm{F390W}}$} &\colhead{F606W}&\colhead{$\sigma_{\rm{F606W}}$}&\colhead{F110W} &\colhead{$\sigma_{\rm{F110W}}$}&\colhead{F160W}  &\colhead{$\sigma_{\rm{F160W}}$} &\colhead{T$_{\rm{eff}}$\tablenotemark{a}} &\colhead{$\sigma_{\rm{T_{\rm{eff}}}}$}\\
\colhead{ } &\colhead{ } &\colhead{(J2000)}&\colhead{(J2000)}  &\colhead{(mag)}  &\colhead{(mag)}  &\colhead{(mag)}  &\colhead{(mag)} &\colhead{(mag)} &\colhead{(mag)} &\colhead{(mag)} &\colhead{(mag)} &\colhead{(K)} &\colhead{(K) }   \\
}\startdata
WD 1  & Swath 1 &00:22:40.912   &  -72:10:32.52 & 25.396 &  0.0351 &   25.362 &  0.0320 &   25.500 &  0.1301 &   25.515 &  0.2238&  12047 &   1602 \\  
WD 2  & Swath 1 &00:22:25.686   &  -72:10:11.84 & 23.620 &  0.0207 &   23.950 &  0.0194 &   24.473 &  0.0464 &   24.430 &  0.0705&  20698 &   1361 \\
WD 3  & Swath 2 &00:21:57.095   &  -72:10:18.59 & 24.263 &  0.0223 &   24.381 &  0.0213 &   24.866 &  0.0623 &   24.898 &  0.1030&  16132 &   1041 \\  
WD 4  & Swath 2 &00:22:00.123   &  -72:10:12.02 & 24.399 &  0.0163 &   24.495 &  0.0182 &   24.722 &  0.0566 &   24.732 &  0.0987&  14570 &    720 \\ 
WD 5  & Swath 2 &00:22:00.719   &  -72:09:15.09 & 25.217 &  0.0270 &   25.125 &  0.0322 &   25.303 &  0.0763 &   25.297 &  0.1365&  11687 &   1161 \\ 
WD 6  & Swath 2 &00:22:26.133   &  -72:09:20.01 & 24.108 &  0.0134 &   24.287 &  0.0235 &   24.633 &  0.0612 &   24.631 &  0.1141&  16372 &    921 \\ 
WD 7  & Swath 2 &00:22:01.506   &  -72:08:34.31 & 23.314 &  0.0220 &   23.684 &  0.0153 &   24.070 &  0.0386 &   24.062 &  0.0683&  20458 &   1281 \\ 
WD 8  & Swath 2 &00:22:25.423   &  -72:09:09.83 & 25.257 &  0.0338 &   25.138 &  0.0316 &   25.325 &  0.0828 &   25.373 &  0.1618&  11566 &   1281 \\ 
WD 9  & Swath 2 &00:22:23.633   &  -72:09:00.06 & 23.728 &  0.0242 &   24.001 &  0.0305 &   24.552 &  0.0550 &   24.795 &  0.1454&  20257 &   1722 \\ 
WD 10 & Swath 3 &00:21:31.487   &  -72:08:35.91 & 23.950 &  0.0213 &   24.129 &  0.0239 &   24.624 &  0.0618 &   24.996 &  0.1379&  17574 &   1201 \\ 
WD 11 & Swath 3 &00:21:39.401   &  -72:09:03.65 & 25.658 &  0.0271 &   25.498 &  0.0292 &   25.538 &  0.1380 &   25.140 &  0.2300&   9684 &   1041 \\ 
WD 12 & Swath 3 &00:21:51.751   &  -72:09:31.15 & 24.491 &  0.0146 &   24.527 &  0.0143 &   24.726 &  0.0558 &   24.953 &  0.1187&  13889 &    640 \\ 
WD 13 & Swath 3 &00:22:01.247   &  -72:09:16.43 & 25.190 &  0.0318 &   25.088 &  0.0218 &   25.168 &  0.0731 &   25.056 &  0.1274&  10966 &    961 \\ 
WD 14 & Swath 3 &00:21:47.243   &  -72:08:22.44 & 25.310 &  0.0232 &   25.168 &  0.0257 &   25.292 &  0.0747 &   24.953 &  0.0945&  10285 &    640 \\ 
WD 15 & Swath 3 &00:21:47.989   &  -72:08:15.89 & 25.613 &  0.0291 &   25.463 &  0.0316 &   25.542 &  0.0982 &   25.239 &  0.1113&  10045 &    801 \\ 
WD 16 & Swath 3 &00:21:41.681   &  -72:07:49.43 & 24.062 &  0.0179 &   24.265 &  0.0182 &   24.778 &  0.0581 &   24.658 &  0.0921&  17374 &   1001 \\ 
WD 17 & Swath 3 &00:21:42.075   &  -72:07:47.08 & 24.640 &  0.0263 &   24.711 &  0.0233 &   25.057 &  0.0772 &   25.144 &  0.1618&  14730 &   1041 \\ 
WD 18 & Swath 3 &00:21:46.162   &  -72:07:54.27 & 23.548 &  0.0210 &   23.877 &  0.0178 &   24.352 &  0.0638 &   24.332 &  0.0914&  19977 &   1401 \\ 
WD 19 & Swath 3 &00:22:02.034   &  -72:08:35.65 & 23.332 &  0.0215 &   23.688 &  0.0191 &   24.209 &  0.0448 &   24.371 &  0.0739&  22220 &   1481 \\ 
WD 20 & Swath 3 &00:21:53.267   &  -72:08:02.38 & 24.810 &  0.0203 &   24.744 &  0.0225 &   25.257 &  0.0831 &   25.381 &  0.2374&  13048 &    961 \\ 
WD 21 & Swath 4 &00:21:30.986   &  -72:08:00.98 & 24.132 &  0.0257 &   24.272 &  0.0247 &   24.622 &  0.0529 &   24.548 &  0.0971&  15812 &   1041 \\ 
WD 22 & Swath 5 &00:21:07.243   &  -72:04:31.54 & 24.833 &  0.0275 &   24.712 &  0.0213 &   25.087 &  0.0823 &   24.955 &  0.1090&  12207 &   1001 \\ 
WD 23 & Swath 5 &00:21:15.803   &  -72:06:00.35 & 24.392 &  0.0309 &   24.438 &  0.0261 &   24.956 &  0.0803 &   24.806 &  0.1638&  14971 &   1201 \\ 
WD 24 & Swath 5 &00:21:17.805   &  -72:04:23.82 & 23.039 &  0.0199 &   23.469 &  0.0151 &   24.030 &  0.0558 &   23.951 &  0.0668&  23221 &   1401 \\ 
WD 25 & Swath 5 &00:21:23.488   &  -72:03:44.75 & 23.803 &  0.0237 &   24.057 &  0.0245 &   24.460 &  0.0606 &   24.654 &  0.0875&  18816 &   1361 \\ 
WD 26 & Swath 5 &00:21:32.257   &  -72:05:38.40 & 22.316 &  0.0209 &   22.940 &  0.0134 &   23.570 &  0.0345 &   23.641 &  0.0564&  31712 &   2563 \\ 
WD 27 & Swath 6 &00:21:26.729   &  -72:01:40.22 & 25.413 &  0.0376 &   25.337 &  0.0198 &   25.266 &  0.0843 &   25.044 &  0.1158&   9924 &    841 \\ 
WD 28 & Swath 6 &00:21:23.115   &  -72:03:44.39 & 23.816 &  0.0156 &   24.088 &  0.0360 &   24.429 &  0.0449 &   24.652 &  0.0810&  18655 &   1121 \\ 
WD 29 & Swath 7 &00:21:39.901   &  -71:59:46.59 & 24.809 &  0.0260 &   24.789 &  0.0226 &   25.096 &  0.0698 &   25.418 &  0.1630&  13569 &   1041 \\ 
WD 30 & Swath 7 &00:21:31.482   &  -72:01:05.77 & 24.259 &  0.0117 &   24.505 &  0.0125 &   24.822 &  0.0581 &   25.031 &  0.0938&  17614 &    720 \\ 
WD 31 & Swath 7 &00:21:48.558   &  -72:00:42.08 & 23.079 &  0.0356 &   23.569 &  0.0165 &   24.071 &  0.0388 &   24.183 &  0.0619&  25544 &   2242 \\ 
WD 32 & Swath 8 &00:21:55.432   &  -71:58:34.17 & 24.546 &  0.0294 &   24.549 &  0.0256 &   24.842 &  0.0576 &   24.642 &  0.0772&  13329 &   1041 \\ 
WD 33 & Swath 8 &00:22:00.315   &  -71:58:30.80 & 25.161 &  0.0322 &   25.080 &  0.0281 &   25.279 &  0.0724 &   24.927 &  0.0880&  11086 &    961 \\ 
WD 34 & Swath 8 &00:21:59.219   &  -71:58:45.40 & 25.135 &  0.0303 &   25.140 &  0.0240 &   25.094 &  0.0649 &   25.171 &  0.1203&  11246 &   1121 \\ 
WD 35 & Swath 8 &00:21:57.370   &  -71:59:11.32 & 22.845 &  0.0233 &   23.291 &  0.0186 &   23.843 &  0.0332 &   23.997 &  0.0509&  25584 &   1642 \\ 
WD 36 & Swath 8 &00:22:05.843   &  -71:59:28.07 & 24.740 &  0.0295 &   24.749 &  0.0277 &   24.944 &  0.0686 &   24.924 &  0.1045&  13209 &   1161 \\ 
WD 37 & Swath 9 &00:22:21.111   &  -71:57:49.36 & 23.525 &  0.0305 &   23.832 &  0.0247 &   24.309 &  0.0454 &   24.645 &  0.0911&  21419 &   1802 \\ 
WD 38 & Swath 9 &00:22:12.785   &  -71:58:54.71 & 25.127 &  0.0260 &   25.015 &  0.0220 &   25.241 &  0.0781 &   25.100 &  0.1078&  11446 &    961 \\ 
WD 39 & Swath 10&00:22:55.003   &  -71:57:27.72 & 24.810 &  0.0188 &   24.749 &  0.0178 &   25.060 &  0.0700 &   25.658 &  0.1799&  12888 &    881 \\ 
WD 40 & Swath 10&00:22:59.788   &  -71:57:41.86 & 25.047 &  0.0274 &   24.976 &  0.0250 &   25.020 &  0.0647 &   24.961 &  0.1094&  11006 &    881 \\ 
WD 41 & Swath 10&00:22:36.414   &  -71:57:47.74 & 24.791 &  0.0306 &   24.775 &  0.0284 &   24.988 &  0.0719 &   25.118 &  0.1012&  13409 &   1161 \\ 
WD 42 & Swath 10&00:22:56.251   &  -71:57:55.80 & 22.920 &  0.0167 &   23.420 &  0.0225 &   23.819 &  0.0340 &   23.777 &  0.0557&  23461 &   1201 \\ 
WD 43 & Swath 10&00:22:33.763   &  -71:58:55.87 & 25.673 &  0.0391 &   25.548 &  0.0274 &   25.462 &  0.0908 &   25.290 &  0.1584&   9804 &    921 \\ 
WD 44 & Swath 11&00:22:59.738   &  -71:57:42.01 & 25.090 &  0.0386 &   25.060 &  0.0247 &   25.160 &  0.0726 &   25.228 &  0.2013&  11927 &   1441 \\ 
WD 45 & Swath 12&00:23:33.344   &  -71:58:26.77 & 24.259 &  0.0177 &   24.387 &  0.0193 &   24.446 &  0.0725 &   24.447 &  0.1015&  14690 &    801 \\ 
WD 46 & Swath 12&00:23:26.967   &  -71:58:57.50 & 23.188 &  0.0194 &   23.617 &  0.0193 &   24.063 &  0.0596 &   24.366 &  0.1228&  23141 &   1561 \\ 
WD 47 & Swath 12&00:23:32.372   &  -71:59:45.70 & 22.430 &  0.0133 &   23.026 &  0.0158 &   23.507 &  0.0376 &   23.823 &  0.0538&  29509 &   1762 \\ 
WD 48 & Swath 12&00:23:21.810   &  -71:59:18.03 & 24.233 &  0.0179 &   24.376 &  0.0250 &   24.915 &  0.0882 &   25.064 &  0.1247&  16653 &   1121 \\ 
WD 49 & Swath 12&00:23:37.390   &  -72:00:36.89 & 23.387 &  0.0151 &   23.753 &  0.0208 &   24.122 &  0.1033 &   24.002 &  0.1056&  20017 &   1361 \\ 
WD 50 & Swath 12&00:23:28.612   &  -72:00:43.42 & 24.186 &  0.0227 &   24.289 &  0.0250 &   24.459 &  0.0823 &   24.690 &  0.1327&  14770 &   1001 \\ 
WD 51 & Stare   &00:23:32.099   &  -72:09:39.30 & 23.982 &  0.0346 &   24.166 &  0.0248 &   24.569 &  0.0471 &   24.632 &  0.2060&  17254 &   1481 \\ 
WD 52 & Stare   &00:23:00.862   &  -72:10:38.40 & 25.293 &  0.0497 &   25.221 &  0.0258 &   25.361 &  0.0681 &   25.014 &  0.0842&  10725 &    921 \\
WD 53 & Stare   &00:23:17.641   &  -72:09:53.81 & 25.215 &  0.0370 &   25.103 &  0.0244 &   25.282 &  0.0706 &   25.734 &  0.2999&  11687 &   1281 \\
WD 54 & Stare   &00:23:09.435   &  -72:10:11.30 & 23.998 &  0.0178 &   24.209 &  0.0172 &   24.527 &  0.0241 &   24.668 &  0.0740&  16973 &    760 \\
WD 55 & Stare   &00:23:18.825   &  -72:09:08.07 & 23.725 &  0.0201 &   24.025 &  0.0175 &   24.597 &  0.0727 &   24.505 &  0.0780&  19416 &   1241 \\
WD 56 & Stare   &00:23:12.136   &  -72:09:23.12 & 25.264 &  0.0520 &   25.184 &  0.0252 &   25.310 &  0.0333 &   25.692 &  0.1589&  12047 &   1081 \\
WD 57 & Stare   &00:23:03.033   &  -72:09:05.57 & 25.514 &  0.0426 &   25.467 &  0.0291 &   25.430 &  0.0524 &   25.235 &  0.1018&  10325 &    720 \\
WD 58 & Stare   &00:23:13.822   &  -72:08:18.48 & 23.385 &  0.0147 &   23.794 &  0.0198 &   24.341 &  0.0291 &   24.294 &  0.0418&  23341 &   1001 \\
WD 59 & Stare   &00:22:55.764   &  -72:08:56.88 & 22.639 &  0.0105 &   23.219 &  0.0159 &   23.758 &  0.0129 &   23.890 &  0.0262&  29028 &    801 \\
\enddata
\tablenotetext{a}{Effective temperature estimates assume a white dwarf mass of $0.53$ M$_{\odot}$.}
\end{deluxetable}
\clearpage


\vspace{1cm}

\figsetstart
\figsetnum{3}
\figsettitle{Images of our WD sample}

\figsetend
\figsetstart
\figsetnum{6}
\figsettitle{Spectral Energy Distributions and likelihood fits for our WDs}

\figsetgrpstart
\figsetgrpnum{3.1}
\figsetgrptitle{WD 1}
\figsetplot{f3_1.eps}
\figsetgrpnote{A four panel image of WD 1. Within the four panels, the {\it upper left} is F390W, {\it upper right} is F606W, {\it lower left} is F110W, and {\it lower right} is F160W.  The WD is centered in the 2\arcsec~x~2\arcsec\ images and its photometry is listed in Table~\ref{tab:wds.}}
\figsetgrpend

\figsetgrpstart
\figsetgrpnum{6.1}
\figsetgrptitle{WD 1}
\figsetplot{f6_1.eps}
\figsetgrpnote{The SED for WD 1.  The {\it filled squares} are our photometric measurements in F390W, F606W, F110W, and F160W with uncertainties.   The wavelength range for each filter is also shown as the equivalent width of the filter divided by the maximum throughput, as defined by the WFC3 instrument handbook.  The {\it red triangles}, connected by the {\it red line} is the best fit model STMAG magnitudes.}
\figsetgrpend

\figsetend

\end{document}